\documentclass[doublecol]{epl2}
\usepackage{amsmath}
\title{Nonlinearity enhanced interfacial thermal conductance and rectification}
\shorttitle{Nonlinear Interfacial Thermal Transport} 

\author{Lifa~Zhang\inst{1} \and Juzar Thingna\inst{1} \and Dahai~He\inst{2} \and Jian-Sheng Wang \inst{1} \and Baowen~Li\inst{1,3,4} }
\shortauthor{L.~Zhang \etal}

\institute{
  \inst{1} Department of Physics and Centre for Computational
Science and Engineering, National University of Singapore, Singapore
117542, Republic of Singapore\\
  \inst{2} Department of Physics and Institute of Theoretical Physics and Astrophysics, Xiamen University, Xiamen 361005, China\\
 \inst{3} NUS Graduate School for Integrative
Sciences and Engineering, Singapore 117456, Republic of Singapore \\
\inst{4}NUS-Tongji Center for Phononics and Thermal Energy Science and
Department of Physics, Tongji University, 200092 Shanghai, PR
China
}

\abstract{
We study the nonlinear interfacial thermal transport across atomic junctions by the quantum self-consistent mean field (QSCMF) theory  based on nonequilibrium Green's function approach; the QSCMF theory we propose is very precise and matches well with the exact results from quantum master equations.  The nonlinearity at the interface is studied by effective temperature dependent interfacial coupling calculated from the QSCMF theory.  We find that nonlinearity can provide an extra channel for phonon transport in addition to the phonon scattering which usually blocks heat transfer.  For weak linearly coupled interface, the nonlinearity can enhance the interfacial thermal transport; with increasing nonlinearity or temperature, the thermal conductance shows nonmonotonical behavior. The interfacial nonlinearity also induces thermal rectification, which depends on the mismatch of the two leads and also the interfacial linear coupling. }

\begin{document}

\maketitle

\section{Introduction}
In modern electronics, due to the rapid increasing power density, accumulation of heat becomes an obstacle  for further progress of microelectronic devices; thus the heat dissipation and manipulation has been recognized to be a crucial issue in information and energy technologies \cite{cahill03}. Especially, as the dimensions of materials shrink into the nanoscale, interfaces dramatically affect the thermal transport \cite{costescu04,hu10,hsieh11,zhang11} making it a lucrative field to explore. At a rough interface the atomic mixing can enhance the thermal transport \cite{tian12}; however, the behavior of the nonlinearity at the interface is not clear, and people do not know whether the nonlinearity can enhance the phonon transport.    Recent progress in  functional thermal devices \cite{rectifiers,transistor,logicgate,memory}, makes  the emerging new field -- phononics very attractive \cite{li12}.  In phononics, the most fundamental property of phononic devices is thermal rectification, which is known to be realized by combining the system inherent anharmonicity with structural asymmetry  \cite{segal05,zhang09}. Whether the interface itself can induce thermal rectification is still an open question; if yes, the property of the interfacial rectification is quite interesting and helpful for both theorists and experimentalists.

To investigate the thermal transport across interface, the most widely applied models are the acoustic mismatch model \cite{little59} and the diffuse mismatch model \cite{swartz89}. Both models offer limited accuracy in nanoscale interfacial resistance predictions \cite{stevens05,reddy05} because they make simple assumptions and neglect atomic details of actual interfaces.  Classical molecular dynamics simulation is another widely used method in phonon transport and has been applied to interfacial thermal transport \cite{li05,choi05,term09,landry09,ong10}; however due to its classical nature, it is not accurate below the Debye temperature and can not capture the quantum effects.  To study the nonlinear (anharmonic) thermal transport, the effective phonon theory has been recently introduced in some dynamical models \cite{li06,li07,hu06}; and the quantum correction one \cite{he08} can be used to study the low temperature thermal transport. Despite their successes, such theories can not be well applied to nonlinear interfacial transport due to the inherent weak system-bath coupling assumption required for the validity of Feynman-Jensen inequality \cite{heunpb}.  Another effective approach, the nonequilibrium Green's function method  which originates from the study of electronic transport \cite{haug96}, has been applied to study the quantum phonon transport \cite{mingo06,lu07,wang08}, phonon Hall effect \cite{zhang09b} and topological magnon insulator \cite{zhang13}.

In this paper, based on the nonequilibrium Green's function method, to avoid the perturbation approximation we develop the QSCMF theory for the nonlinear thermal transport, which can be applied to thermal transport in an arbitrary strength nonlinear interface. Then we study the interfacial thermal transport for a model as shown in Fig.\ref{fig1}(a); thermal conductance and rectification across the interface are studied with an effective temperature-dependent-harmonic interfacial coupling calculated form the QSCMF theory.

\section{Model and Hamiltonian}
\begin{figure}
\includegraphics[width=3.4 in]{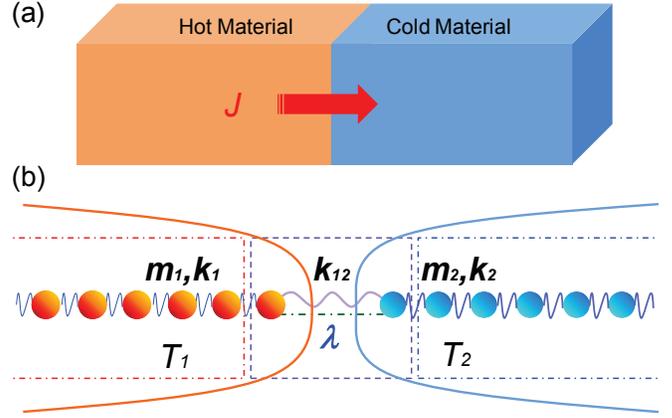}
\caption{\label{fig1}(Color online) (a) Heat transport in a solid-solid interface. The arrow shows the heat transport from the hot side to the cold side. (b) The atomic junction model of the solid-solid interface. The solid line regions are two semi-infinite atomic chains which are coupled by a harmonic spring with strength $k_{12}$. In addition to which, the two regions also have a fourth order nonlinear coupling  $\lambda$. For the two semi-infinite chains, the mass and spring constant are $m_1$, and $k_1$, $m_2$, and $k_2$, respectively.  The interface model can be partitioned to three parts, the center (dashed line) and the leads (dash-dotted lines) with temperatures $T_1$ and $T_2$. }
\end{figure}

We study the interfacial thermal transport with nonlinear coupling at the solid-solid interface as shown in Fig.~\ref{fig1}(a). To manifest the effect of the interface we exclude the nonlinear phonon transport in the two materials, and there is no disorder nor defect in the whole system,  thus the only thermal resistance comes from the interface. Such model really uncovers the thermal transport properties of the interface. To study the longitudinal transport, that is, the cross-plane interfacial transport, we simplify the problem further to one dimensional interface as shown in Fig.~\ref{fig1}(b), where two linear semi-infinite atomic chains (solid-line regimes) connect each other by the interfacial linear coupling $k_{12}$ and nonlinear interaction $\lambda$, which is similar to the one-dimensional interfacial model in Ref.\cite{zhang11,hu10} where there is only linear coupling at the interface. The Hamiltonian of the total system is

\begin{equation}\label{eq_ham}
H = H_1  + H_2  + \frac{1}{2}k_{12} (x_{1,1}  - x_{2,1} )^2  + \frac{1}{4}\lambda (x_{1,1}  - x_{2,1} )^4,
\end{equation}
with
\begin{equation}
H_{\alpha}  = \sum\limits_{i = 1}^{N_\alpha  } {\frac{1}{2}m_\alpha  \dot x_{\alpha ,i}^2 }  + \sum\limits_{i = 1}^{N_\alpha   - 1} {\frac{1}{2}k_\alpha  (x_{\alpha ,i}  - x_{\alpha ,i + 1} )^2 },
\end{equation}
where $N_\alpha\rightarrow\infty$, $\alpha  = 1,2$.  In our model, the scattering for phonons only comes from the interface while the phonon transport in the two semi-infinite leads is ballistic. Thus we can partition the system into three parts ($L$, $C$, $R$), where the atoms at the interface are regarded as center (dashed-line part)  and leads $L$ and $R$ (dash-dotted-line regimes) are harmonic as shown in Fig.\ref{fig1} (b). Using a mass-normalized displacement $u_j  = \sqrt {m_j } x_{j,1} $, the center Hamiltonian can be written as
\begin{equation}\label{eq_hamc}
H_C  = \frac{1}{2}\dot U_c^T \dot U_c  + \frac{1}{2}U_c^T K^C U_c  + \frac{1}{4}\sum\limits_{i,j,k,l = 1}^2 {T_{ijkl} } u_i u_j u_k u_l,
\end{equation}
where $U_c=(u_1, u_2)^T$, $T_{ijkl}=(-1)^{i+j+k+l}\lambda$ and
\begin{equation}
K^C  = \left( {\begin{array}{*{20}c}
   \frac{k_1  + k_{12}}{m_1}  &  \frac{- k_{12}}{\sqrt {m_1 m_2 } }  \\
   \frac{ - k_{12}}{\sqrt {m_1 m_2 }} & \frac{k_{12}  + k_2}{m_1}   \\
\end{array}} \right).
\end{equation}
For the leads, the Hamiltonian is written as $H_\alpha   = \frac{1}{2}\dot U_\alpha ^T \dot U_\alpha   + \frac{1}{2}U_\alpha ^T K^\alpha  U_\alpha$ with its coupling to the center $H_{\alpha C}=U_\alpha^T V^{\alpha C} U_C$, $\alpha=L,R$. Here the center-lead coupling is the same as the inter-atomic spring constant in the corresponding bath.

\section{Quantum Self-Consistent Mean Field Theory}
We discuss the QSCMF based on NEGF method for a general system where the center hamiltonian has a fourth-order nonlinear interaction as given in Eq.~(\ref{eq_hamc}).  The equation of motion of Green's function \cite{wang08},  without the nonlinearity, is
{\small $
(\frac{\partial ^2 }{{\partial \tau ^2 }} + K^C )G_0 (\tau \tau ') = - I\delta (\tau  - \tau ')- \int d\tau '' \Sigma (\tau \tau '')G_0 (\tau ''\tau '),
$}
where $\Sigma (\tau\tau '')$ is the self energy due to the center-lead coupling. With the nonlinearity, the full Green's function has the equation of motion as
{\small
\begin{equation}\begin{split}
\frac{{\partial ^2 }}{{\partial \tau ^2 }}G_{im} (\tau \tau ') + \sum\limits_j {K_{ij}^C } G_{jm} (\tau \tau ')+ \sum\limits_{jkl} {T_{ijkl} G_{jklm} (\tau \tau \tau \tau ')}  \\   =  - \delta (\tau  - \tau ')\delta _{im} - \sum\limits_j {\int d\tau '' \Sigma _{ij} (\tau \tau '')G_{jm} (\tau ''\tau ')}
\end{split}\end{equation} }
where $G_{jklm} (\tau_j \tau_k \tau_l \tau_m )$ is the four-point Green's function. Under the self-consistent mean field approximation, the four-point Green's function can be represented by the two-point Green's function, and we have
{\small
$
 -i \frac{1}{\hbar}G(\tau _j \tau _k \tau _l \tau _m ) =  G(\tau _j \tau _k )G(\tau _l \tau _m ) + G(\tau _j \tau _l )G(\tau _k \tau _m ) + G(\tau _j \tau _m )G(\tau _k \tau _l ).
$}
 Thus the full Green's function satisfies
{\small
\begin{equation}\begin{split}
\frac{{\partial ^2 }}{{\partial \tau ^2 }}G_{im} (\tau \tau ') + \sum\limits_j {K_{ij}^C } G_{jm} (\tau \tau ') \\ + 3i\hbar \sum\limits_{jkl} {T_{ijkl} G_{kl} (0)G_{jm} (\tau \tau ')} + \sum\limits_j {\int d\tau '' {\Sigma _{ij} (\tau \tau '')G_{jm} (\tau ''\tau ')} }\\ =  - \delta (\tau  - \tau ')\delta _{im},
\end{split}\end{equation} }
by using the symmetry of $T_{ijkl}$ with respect to the permutation of the indices. Thus we introduce an effective dynamic matrix
\begin{equation}\label{eq_effkc}
\tilde K_{ij}^C  = K_{ij}^C  + 3i\hbar \sum\limits_{kl} {T_{ijkl} G_{kl}(0)} = K_{ij}^C  + 3\sum\limits_{kl} {T_{ijkl} \left\langle {u_k u_l } \right\rangle }.
\end{equation}
The nonlinearity only has the effect to modulate the dynamic matrix. It is important to note that the results are independent of the partition size of the center due to $T_{ijkl}=0$ every where else except the two atoms at the interface.

Equation (\ref{eq_effkc}) together with
$
G^{r} = [(\omega+i\eta)^{2}-\tilde K_{ij}^C-\Sigma^{r}]^{-1}, \;
G^<=G^r \Sigma^< G^a, \;
\left\langle {u_k u_l } \right\rangle  = i\hbar  G_{kl}^ <  (t = 0) =2 i\hbar
\int_0^\infty  {G_{kl}^ <  (\omega )d\omega /(2\pi )}
$
can be self-consistently calculated. Since the problem is now effectively harmonic, the heat current still satisfies the Landauer formula $
J= \int_0^\infty  {\frac{{d\omega }}{{2\pi }}\hbar \omega T[\omega ](f_L  - f_R )} $, $f_\alpha=1/(e^{\hbar\omega/(k_BT_\alpha)}-1)$, and thermal conductance is defined as $\sigma= |J/(T_L-T_R)|$, while the transmission $T[\omega ] = {\rm Tr}( G^r \Gamma _L G^a \Gamma _R )$ is temperature dependent. In later calculation, we will set $\hbar=1, k_B=1$ for simplicity. For the conversion from the dimensionless unit to physical units, we take energy unit $[E]=1$ mev, length unit is $[L]=1 {\rm \AA}$, then temperature unit $[T]$ is 11.6 K, thermal conductance $[\sigma]=20.9 {\rm nW/mK}$, the spring constant unit  $[k]=1$ mev/${\rm \AA}^2$ and nonlinearity unit $[\lambda]=1$ mev/${\rm \AA}^4$.

Applying our QSCMF theory to the nonlinear interface problem of Eq.~(\ref{eq_ham}), we find that the nonlinearity plays a role to modulate the interfacial linear coupling $k_{12}$, the effective one is
\begin{equation}\label{eq_effk12}
  k_{12\rm{eff}}  = k_{12}  + 3\lambda \bigl(\frac{{\left\langle {u_1^2 } \right\rangle }}{{m_1 }} - 2\frac{{\left\langle {u_1 u_2 } \right\rangle }}{{\sqrt {m_i m_j } }}  + \frac{{\left\langle {u_2^2 } \right\rangle }}{{m_2 }}\bigr).
\end{equation}
Thus it is clear that all the scattering occurs only at the interface, since all other parts are harmonic.

\begin{figure}[t]
\includegraphics[width=3.40 in,height=2.2 in, angle=0]{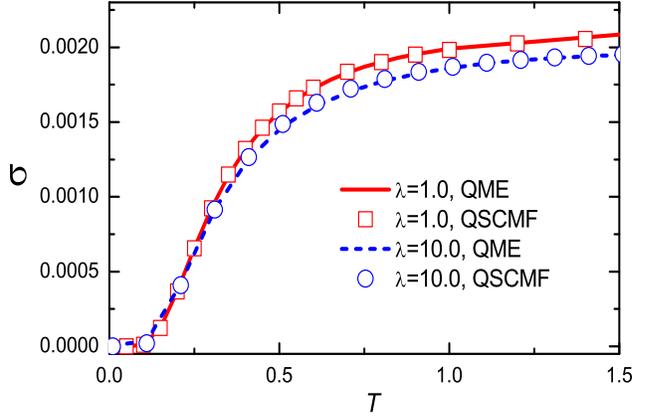}
\caption{ \label{fig2} (Color online) Comparison of the results for thermal conductance of the interface as illustrated in Fig.~\ref{fig1}(b) between QSCMF theory and quantum master equation method.  We add a small onsite potential $0.01$ to each atom in both leads.  $m_1=m_2=1.0$, $k_1=k_2=1.0$, and $K_{11}^C=1.5,K_{22}^C=1.5$, $K_{12}^C=-0.5,K_{21}^C=-0.5$. The system-bath coupling is $0.1$ .}
\end{figure}
\section{Comparison with the Quantum Master Equation}
While the effective phonon theory and its quantum correction one are valid in the linear response region at weak system-bath coupling, our QSCMF theory can study the nonequilibrium thermal transport under larger temperature bias at any system-bath coupling.  For anharmonic systems with arbitrary strength of anharmonicity
under the weak system-bath coupling approximation, Redfield quantum master equation is very suited solution to study the thermal transport \cite{segal05,thingna12a}.
In our interface problem, we partition the atoms in the interface as a center, and the center-lead coupling is the same as the inter-atomic spring constant in the corresponding bath. Thus for such system, the system-bath coupling is always strong such that the quantum master equation cannot be applied which is limited in the weak system-bath coupling limit. However, the QSCMF method can be applied.

We numerically compare our theory with the quantum master equation at a weak system-bath coupling, as shown in Fig.~\ref{fig2}. In Fig.~\ref{fig2}, the system-bath coupling is $0.1$ while the inter-atomic spring constant in the leads are $1.0$ such that the weak-system-bath-coupling condition is well satisfied. A small onsite potential $0.01$  is added to each atom in both leads to avoid the divergency in quantum master equation method. From Fig.~\ref{fig2} we find that the results from QSCMF perfectly matches those from the quantum master equation method for different nonlinearity. Therefore the QSCMF method is verified by the quantum master equation method at weak system-bath coupling; however, the QSCMF is not limited in this weak coupling, which can be applied to arbitrary-system-bath-coupling systems.

\section{ Numerical Results on Interface thermal Transport}
\subsection {Nonlinearity Suppressed Thermal Transport in Homogenous Systems}
\begin{figure}[t]
\includegraphics[width=3.40 in,height=2.2 in,angle=0]{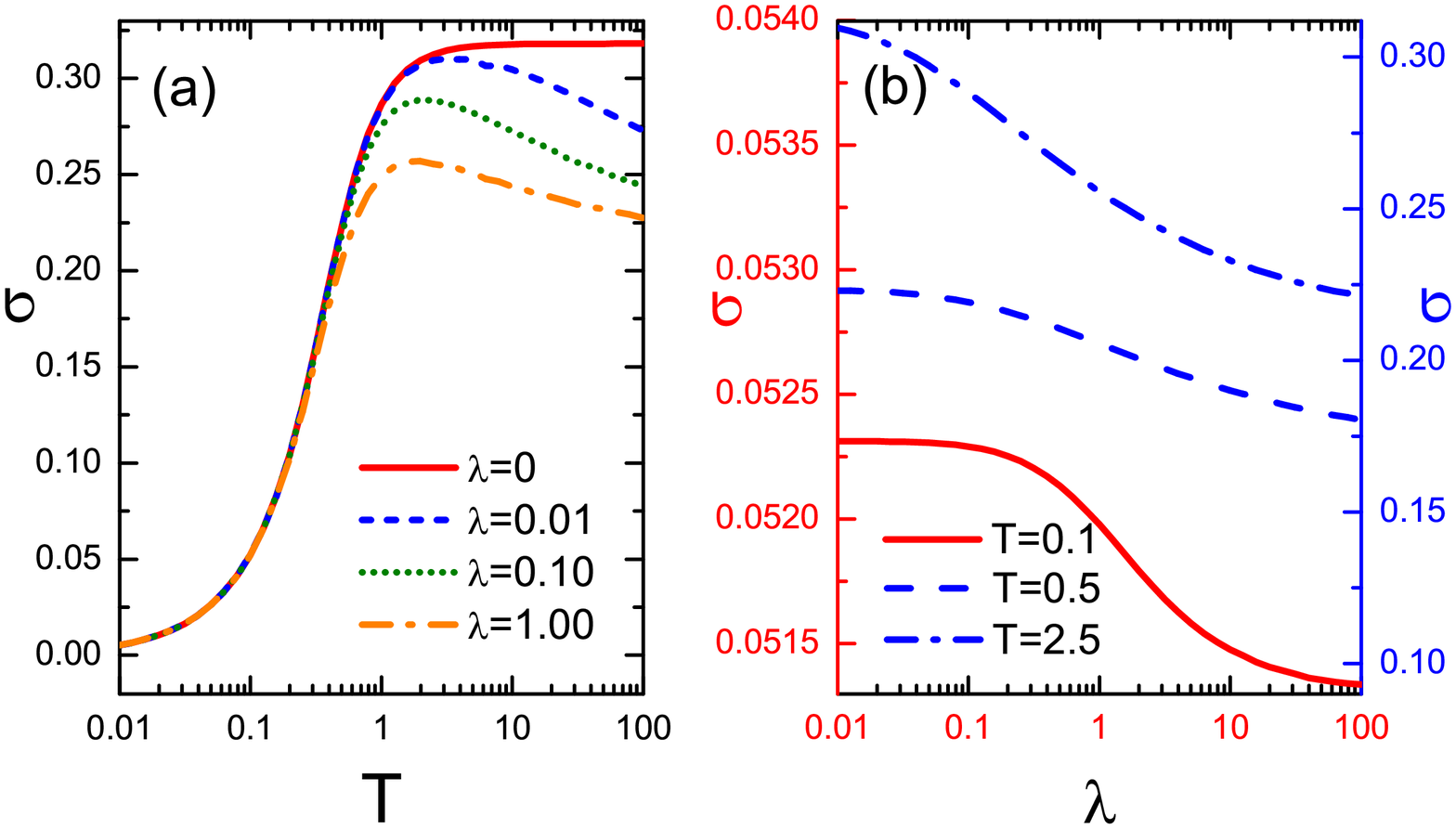}
\caption{ \label{fig3} (Color online) Interfacial thermal conductance as a function of $T$ ($T=(T_L+T_R)/2$)(a) and $\lambda$ (b). $m_1=m_2=1.0$, $k_1=k_2=1.0$ and $k_{12}=1.0$. In (b), the solid line ($T=0.1$) corresponds to left scale, and the dashed line ($T=0.5$) and dotted line ($T=2.5$) correspond to right scale.}
\end{figure}

Using the QSCMF theory we proposed above, the interface nonlinearity can be studied for the interfacial thermal transport. For the homogeneous lattice with $k_1=k_2=k_{12}$, we calculate the interfacial thermal conductance for different nonlinearity $\lambda$ as shown in Fig.~\ref{fig3}(a). With zero nonlinearity, the thermal conductance increases with the temperature increasing, and tends to a constant due to the saturate of phonon modes contributing to the thermal transport.  However, with nonzero interfacial nonlinearity the thermal conductance decreases at high temperatures due to the dominant  scattering coming from the nonlinear interface coupling. In the low-temperature regime, the thermal conductance almost coincides with the ballistic transport and the nonlinearity almost has no effect on thermal transport. With increasing temperature, more phonon modes are excited which wins the suppressing effect nonlinear scattering such that the conductance increases. At certain temperature, the conductance arrives its maximum, after which it will decrease since the nonlinear scattering effect which defeats the enhancement effect from more excited phonon modes.  With increasing nonlinearity the thermal conductance decreases due to the larger phonon scattering at the interface. As shown in Fig.~\ref{fig3}(b), the nonlinearity always decreases the thermal transport for the homogeneous systems. The larger interfacial nonlinearity makes the system more nonhomogeneous such to induce more scattering to the phonon transport.

\subsection {Nonlinearity Enhanced Interfacial Thermal Conductance}
\begin{figure}[t]
\includegraphics[width=3.40 in,height=2.2 in,angle=0]{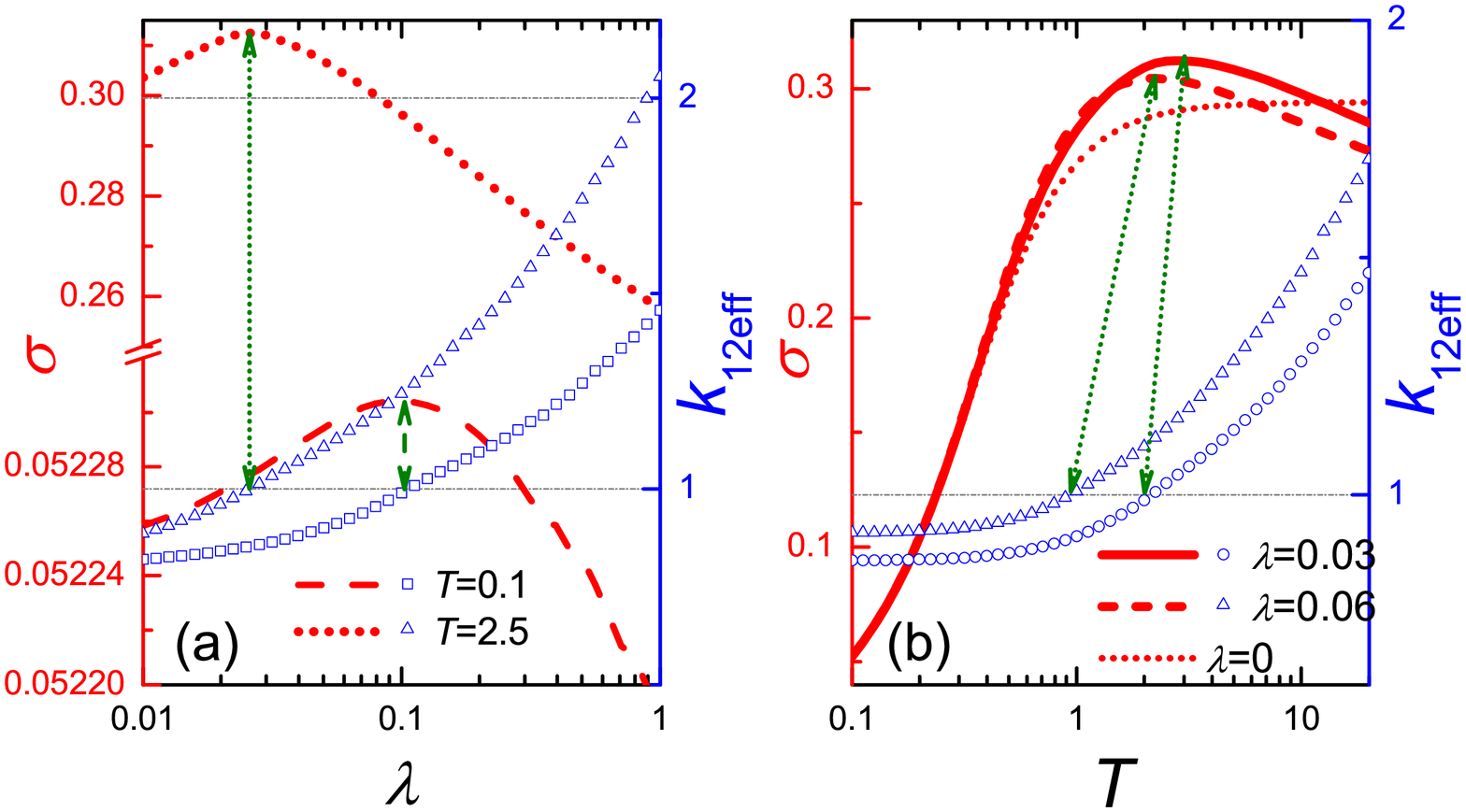}%
\caption{ \label{fig4} (Color online) Interfacial thermal conductance and effective interfacial coupling as functions of $\lambda$ (a) and $T$ (b). $m_1=m_2=1.0$, $k_1=k_2=1.0$ and $k_{12}=0.80$. For both (a) and (b), the solid, dashed and dotted lines correspond to left scale (interfacial thermal conductance), and the scatter points (square, circle and triangle symbols) correspond to right scale (effective interfacial coupling). The lines with two arrows are drawn to guide the eyes.}
\end{figure}

In the weak interfacial coupling regime, that is, $k_{12}<k_1=k_2$, we find that the nonlinear interaction at interface can enhance the thermal transport as shown in Fig~\ref{fig4}. With increasing nonlinearity, the interfacial conductance increases first; after certain maximum the conductance will decrease, as shown in Fig~\ref{fig4}(a).  The maximum conductance coincides with the point where the effective coupling equals to $1$ and the whole system is homogeneous. If $\tilde k_{12}$ increases further the conductance decreases due to the larger scattering at interface. At a fixed linear interfacial coupling $k_{12}<1$, the interfacial nonlinearity makes the effective one $\tilde k_{12}$ larger than $k_{12}$. A larger $\tilde k_{12}$ reduces the difference between the interface and the leads, thus decreases the phonon scattering and allows more phonons to transmit through the interface. Therefore the nonlinearity introduces an extra channel to transport phonons, which enhances the thermal transport. In Fig~\ref{fig4} (b), the maximum of thermal conductance does not coincide with the place of  $ \tilde k_{12}=1$, this is mainly because that the increase in temperature causes more phonon modes to transport so as to delay the maximum of conductance. For a larger nonlinearity, the maximum of thermal conductance is delayed further.

\begin{figure}[t]
\includegraphics[width=3.40 in, height=2.2 in, angle=0]{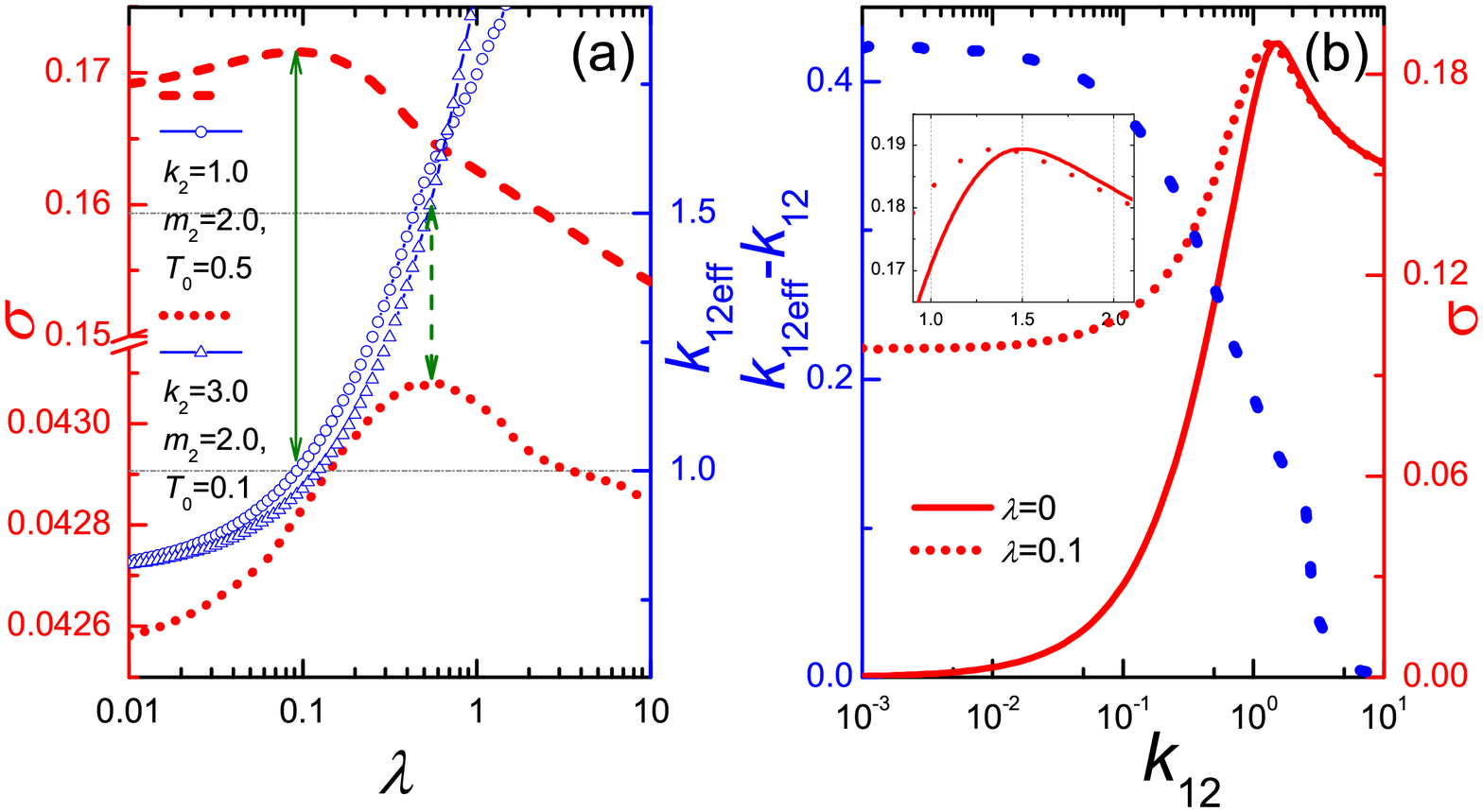}
\caption{ \label{fig5} (Color online) (a) Thermal conductance (left scale) and effective interfacial coupling (right scale) as functions of $\lambda$.  $k_1=m_1=1.0$, $k_{12}=0.8$,  the dashed and dotted lines correspond to interfacial thermal conductance, and the scatter pionts (circle or triangle symbols) correspond to effective interfacial coupling; the lines with two arrows are drawn to guide the eyes.  (b) The effective coupling from the nonlinearity $k_{12\rm{eff}}-k_{12}$ and interfacial thermal conductance $\sigma$ as functions of interfacial linear coupling. The dashed line corresponds to the left scale ($k_{12\rm{eff}}-k_{12}$) and solid and dotted lines correspond to right scale ($\sigma$).  $k_1=m_1=1.0$, $k_2=3.0, m_2=2.0$, $T_0=0.5$; the inset is the zoom-in plot around $k_{12}=1.5$ }
\end{figure}

For a hetero-junction, that is, the two mismatched leads, the thermal conductance can also be enhanced by interfacial nonlinearity. We know that the thermal conductance reaches a maximum when the interfacial coupling equals the harmonic average of the spring constants of the two semi-infinite chains~\cite{zhang11}. As shown in Fig.~\ref{fig5}(a) the conductance of two general hetero-junctions gets to the maximum when the effective coupling $k_{12\rm{eff}}$ equals to $k_{12m}=2k_1k_2/(k_1+k_2)$; after which the thermal conductance decreases. The interfacial conductance can be enhanced in a wide nonlinearity range, as shown in Fig.~\ref{fig5}(a). The enhancement of the conductance due to the nonlinearity depends on the interfacial linear coupling. In order to show this, we plot the thermal conductance as a function of the linear coupling $k_{12}$ in Fig.~\ref{fig5}(b). For a fixed nonlinearity, the thermal conductance enhances largest at zero $k_{12}$, where the only channel to transport phonon comes from the nonlinearity. With increasing $k_{12}$ the conductance increases, but the effective coupling due to the nonlinearity $k_{12\rm{eff}}-k_{12}$ decreases. This effect causes the overall enhancement to decrease.  The inset of Fig.~\ref{fig5}(b) shows that the nonlinear interfacial conductance equals the linear thermal case just before $k_{12}=k_{12m}=1.5$, where the scattering from nonlinearity cancels the contribution from the extra channel introduced by the nonlinearity. At $k_{12}=k_{12m}=1.5$, the linear conductance maximizes and the nonlinearity only adds to the phonon scattering, thus the conductance of nonlinear interface is small. If $k_{12}>>\lambda$, the effective coupling from the nonlinearity $k_{12\rm{eff}}-k_{12}$ nearly vanishes making the interfacial nonlinearity not important anymore, thus the conductance is almost the same with the ballistic one.

\subsection {Nonlinearity Induced Interfacial Thermal Rectification}
From our QSCMF approach, the effective interfacial coupling is temperature dependent; if we reverse the two temperatures of the leads it would be different so that we can observe thermal rectification. If the interface coupling is linear, then Landauer formula applies and reverse temperature only changes the sign of the heat current, thus there is no rectification.  The rectification is defined as \cite{zhang10}:
\begin{equation}
 R =(J_+ - J_-)/ {\rm max }\{J_+, J_-\},
\end{equation}
where $J_+$ is the-forward direction heat flux when $T_L=T_h, T_R=T_c$, and $J_-$ is that of the backward direction when $T_L=T_c, T_R=T_h$. Here $T_h$ and $T_c$ correspond to the temperatures of the hot and cold baths, respectively.

\begin{figure}[t]
\includegraphics[width=3.40 in, height=2.2 in,  angle=0]{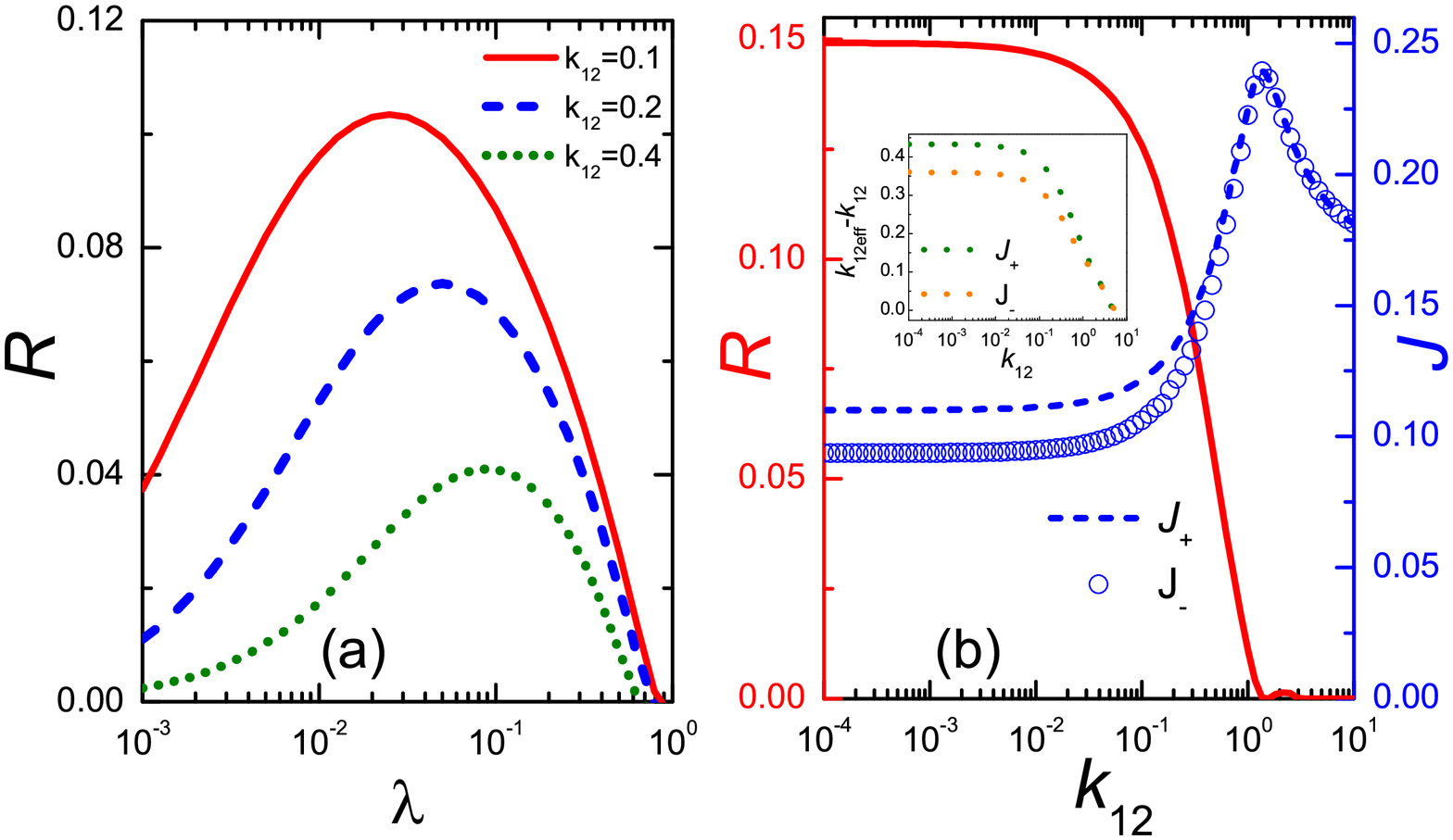}
\caption{ \label{fig6} (Color online) (a) Interfacial thermal rectification as functions of $\lambda$ for different $k_{12}$. (b) Interfacial thermal rectification (left scale) and the heat current (right scale) as functions of $k_{12}$. For both (a) and (b), $k_1=m_1=1.0$, $k_2=3.0, m_2=2.0$, and $T_+=T_0(1+\Delta)$, $T_-=T_0(1-\Delta)$ with $T_0=0.5$, $\Delta=0.5$; for (b) $\lambda=0.1$. The inset in (b): the effective coupling from the nonlinearity $k_{12\rm{eff}}-k_{12}$ as a function of interfacial linear coupling $k_{12}$. }
\end{figure}

With the asymmetric structure the nonlinear interface shows rectification, which depends on the interfacial nonlinearity $\lambda$ and the linear coupling $k_{12}$ as shown in Fig.~\ref{fig6}. If the nonlinearity is zero, there is no rectification. With increasing the nonlinearity, the effective coupling will be different in the forward and backward transport which causes the rectification to increase. When the nonlinearity increases further, the effective coupling will monotonically increase. Therefore the scattering from the interface will play an important role to decrease the flux difference between the forward and backward ones, which causes the rectification to decrease as shown in Fig.~\ref{fig6}(a). With larger linear interfacial coupling, the contribution to the phonon transport from the channel of $k_{12}$ is larger as compared to the extra channel provided by the nonlinearity $\lambda$. Therefore the relative nonlinear effect is weakened and the rectification is reduced as shown in Fig.~\ref{fig6}(b). In the opposite limit, if $k_{12}<<\lambda$ the rectification almost keeps fixed, when the nonlinearity dominates the thermal transport and the difference of the effective coupling between the forward and backward flow has almost no changes. When $k_{12}\sim\lambda$, the rectification decreases fast due to the fast decrease of the value of $k_{12\rm{eff}}-k_{12}$.   If $k_{12}>>\lambda$, the rectification decreases to almost zero since the effective coupling from the nonlinearity $k_{12\rm{eff}}-k_{12}$ nearly vanishes making the nonlinearity not important anymore.

\section{Discussion}
We use a simplified one-dimensional atomic model to investigate the underlying physics in the interfacial thermal transport; based on which we find that the nonlinearity can enhance the thermal transport and the nonlinearity at interface can induce the interfacial thermal rectification. For a general chemical bonding strength across interface, we could represent it as a linear coupling part and nonlinear coupling part; therefore our model can be applied. However, to compare with experimental results on real materials, the simplified model need to be generalized to two-dimensional or three dimensional with input of the real parameters.

The QSCMF theory is a kind of mean-field approximation based on the nonequilibrium Green's function, which is similar as the mean field approach in the literatures \cite{haug96,bruus04}. The QSCMF is a good candidate to solve the nonlinear problem with arbitrary nonlinearity and arbitrary system-bath coupling. By comparing to the quantum master equation which is limited by weak-system-bath coupling, we find the QSCMF is a quite accurate method while it can be applied to strong system-bath coupling. For the interface problem with two-layer atoms, the QSCMF has a very high accuracy in the wide range of temperature and nonlinearity. However, if more layer-atoms includes, the accuracy decreases. If the QSCMF is generalized to two or three dimensional interface with more atoms in the interface, the self-consistent process is time consuming, and we also need to pay more attention to its convergency. The QSCMF is quite good for two-layer-atoms interface which is a reasonable approximation for short-range inter-atom coupling; however for long-range interaction, more layers need to be considered at the interface and the numerical calculation on the nonlinear thermal transport is challenging.
\section{Conclusion}
Based on the NEGF approach, an efficient QSCMF theory is developed to study the nonlinear interfacial thermal transport. We find that the nonlinearity can enhance the interfacial thermal transport at weak linear interfacial coupling while the enhancement vanishes in the strong linear coupling regime. The enhancement can exist at large nonlinearity where the effective coupling is less than the harmonic average of the spring constants of the two semi-infinite chains. Although the leads are linear, the interfacial nonlinearity  can induce rectification provided that the two leads are asymmetric. With increasing the nonlinear coupling, the rectification first increases then decreases. The interfacial rectification also depends on the interfacial linear coupling; it vanishes if the linear interfacial coupling increases far beyond the nonlinear coupling.

\section{Acknowledgements}
L. Z. and B. L. are supported by the grant
R-144-000-300-112 from Ministry of Education of Republic of Singapore.
J.-S. W. and J. T. acknowledge support from a faculty research grant R-144-000-257-112 of NUS. D. H. surported by NSFC (Grant No. 11047185) and FRFCU (Grant No. 2010121009) of China.

\end{document}